\documentclass{aa}

\usepackage{graphicx}
\usepackage{txfonts}
\usepackage{amsmath}
\usepackage{xcolor}
\usepackage[normalem]{ulem}

\bibpunct{(}{)}{;}{a}{}{,}
\begin{document}

   \title{Observing boson stars in binary systems: The case of Gaia BH1}

   \author{Pedro~Passos\inst{1}
          \and
          Héctor~R.~Olivares-Sánchez\inst{2}
          \and
          José A.~Font\inst{3,4}
          \and
          António~Onofre\inst{1}
          }
   
   \institute{Centro de F\'{\i}sica das Universidades do Minho e do Porto (CF-UM-UP), Universidade do Minho, 4710-057 Braga, Portugal\\
              \email{pedropassos199860@gmail.com}\\
              \email{antonio.onofre@cern.ch}
         \and
            Departamento de Matematica da Universidade de Aveiro and Centre for Research and Development Mathematics and Applications (CIDMA), Campus de Santiago, 3810-193 Aveiro, Portugal \\
             \email{h.sanchez@ua.pt}
         \and
            Departamento de Astronom\'{\i}a y Astrof\'{\i}sica, Universitat de Val\`encia, Dr. Moliner 50, 46100, Burjassot (Val\`encia), Spain 
            \and
            Observatori Astronòmic, Universitat de València, C/ Catedrático José Beltrán 2, 46980, Paterna (València), Spain \\
             \email{j.antonio.font@uv.es}
             }
   
   \date{Received ; accepted}

\abstract
{The Gaia experiment recently reported the observation of a binary system composed of a Sun-like star orbiting a dark compact object, known as Gaia BH1. The nature of the compact object remains uncertain. While the Gaia mission identifies it as a black hole candidate, the absence of X-ray or radio detections challenges that interpretation, and alternative exotic compact objects such as boson stars have also been suggested.}
{In this paper, we study whether a boson star could account for the observed properties of the source.} 
{To do so we compute the X-ray luminosity of the central dark object as a result of spherically symmetric (Bondi-Michel) accretion of matter, comparing our results for the cases in which the dark object is a Schwarzschild black hole or a non rotating boson star. Our model incorporates realistic interstellar medium properties, ranging from hot ionized gas to dense molecular clouds. By solving the governing 
equations numerically, we calculate mass accretion rates and derive the resulting Bremsstrahlung X-ray luminosities.}
{Black holes and boson stars fundamentally differ by the absence of an event horizon in the latter, which directly impacts accretion dynamics as there is an accumulation of mass in regions closer to the boson star, which will significantly change the observed X-ray emission. For the Gaia BH1 system we find that accretion onto a black hole yields luminosities of $\sim10^{27} \ \text{erg}\, \text{cm}^{-2}\, \text{s}^{-1}$ which corresponds to an X-ray flux undetectable by Chandra’s sensitivity. On the other hand, boson star accretion can produce observable luminosities in the order of $10^{27} \ \text{to} \ 10^{41} \ \text{erg}\, \text{cm}^{-2}\, \text{s}^{-1}$.
}
{We argue that X-ray observations could help in discriminating the nature of the dark companions in binary systems like Gaia BH1 and test the possibility of boson stars as alternatives to black holes.}

   \keywords{Bondi-Michel Accretion - Boson Stars Physics - Black Hole physics - GAIA BH1 - Chandra X-ray Telescope
               }
  
   \maketitle
   
%

\section{Introduction}


The Gaia mission \citep{gaia_collaboration_gaia_2016} launched by the European Space Agency (ESA) in 2013 is designed to create the most stringent 3D map of our galaxy. Its observations are designed to study and understand the evolutionary conditions and the orbital dynamics of multiple star systems. The importance of these studies allows us to test stellar evolution models, measure stellar masses, and explore the gravitational interactions of diverse systems in different astrophysical environments. Continuous improvement in sensitivity, calibration, and analysis techniques in the astrometric mission allowed Gaia to release its DR3 catalogue \citep{gaia_collaboration_gaia_2023}, which contains more than $8\times 10^5$ data on non-single star dynamics. 
In this catalogue, Gaia reported a peculiar binary system composed of a Sun-like star orbiting a central dark object \citep{2023MNRAS.518.1057E}, referred to as Gaia BH1.  The orbital characteristics of this system, with an orbital period of approximately $185.6$ days and an eccentricity of about $0.45$, diverge from existing knowledge regarding stellar-mass black hole binaries, suggesting a unique evolutionary scenario. The luminous star was observed using spectroscopic methods, allowing researchers to analyze its spectrum and determine its effective temperature ($\approx5860\ \text{K}$), surface gravity, and metallicity. These measurements helped confirm the star's classification as a main-sequence G-dwarf star. In~\cite{2023MNRAS.518.1057E}, the dark object was identified as a black hole candidate via the joint modeling of radial velocities and astrometry, constraining its mass to $9.62\pm 0.18M_{\odot}$. However, the lack of X-ray and radio-wave observations of the central object challenges its unambiguous classification as a black hole.
As a result, alternative proposals have been put forward.  

One such possibility is a boson star, a stable, self-gravitating configuration of bosonic particles -- a macroscopic Bose-Einstein condensate . 
This possibility was proposed by~\cite{pombo_sun-like_2023}, who provided a compelling explanation for the binary system’s dynamics under the boson star assumption for the central dark object. Boson stars are ``exotic'' alternatives to the standard astrophysical compact objects (black holes and neutron stars) constructed from a complex scalar Klein-Gordon field minimally coupled to Einstein's gravity. This framework allows to build theoretical stellar-like solutions in the form of coherent, soliton-like configurations in general relativity (see~\cite{visinelli_boson_2021,liebling_dynamical_2012,2024arXiv240604901B} for recent reviews). The stability of these stars arises from the equilibrium between the gravitational force and the dispersion of the fields. 
Although Derrick's theorem prevents the existence of stable solutions for regular, static, and non-topological fields in flat spacetime \citep{derrick_comments_1964,carloni_derricks_2019}, this can be circumvented 
by adopting a harmonic ansatz for the field while maintaining the time dependence yet keeping a static spacetime metric due to the harmonic nature of the phase. 
Gravitationally, boson stars can imitate black holes in many respects. They can mimic the shadow of black holes \citep{herdeiro_imitation_2021,rosa_shadows_2022, olivares_how_2020}, the spectrum of thin accretion disks \citep{guzman_accretion_2006}, they are feasible alternatives for the evolutionary dynamics in binaries \citep{pombo_sun-like_2023}, and have even been invoked as  alternative explanations to interpret some of the gravitational wave events reported by the LIGO-Virgo-KAGRA collaboration \citep{bustillo_gw190521_2021,bustillo_searching_2023}. 

The study of matter accretion onto a central compact object can shed some light on the nature of the object. Numerical simulations  of fluid accretion onto a boson star in general relativity were first presented in~\cite{meliani_gr-amrvac_2016} where it was shown that accretion can help differentiate boson stars from black holes. Following this approach,  we investigate in this paper whether a boson star could account for the observed properties of the central compact object of Gaia BH1. To do so, we model the accretion dynamics using the Bondi-Michel theory~\citep{Bondi:1952,Michel:1972}, which describes steady, spherically symmetric accretion onto compact objects in general relativity. Bondi's pioneer solution~\citep{Bondi:1952}, limited to Newtonian gravity, introduced critical accretion radii, marking the transition between subsonic and supersonic flow velocities. It  served as the foundation for subsequent modifications and studies, leading the groundwork for many astrophysical applications. Building on Bondi's framework, \cite{Michel:1972} extended the theory into the relativistic regime by providing an exact solution for the spherical accretion of a polytropic gas onto a Schwarzschild black hole. Presently, models encompass various equations of state, from non-relativistic to ultra-relativistic regimes (see~\cite{chaverra_michel_2016}). Further developments have enhanced those models by incorporating the self-gravity of the fluid~\citep{malec_fluid_1999,lora-clavijo_pbh_2013}, low angular momentum effects \citep{mach_relativistic_2018}, magnetic fields \citep{ressler_magnetically_2021}, and radiation \citep{weih_two-moment_2020}. Bondi's model has also been extended to Kerr black holes, where angular momentum induces axisymmetric accretion solutions, accompanied by shock waves \citep{chakrabarti_solutions_1996}.

The Bondi-Michel accretion framework is applied here to describe matter inflow for two different compact objects, a Schwarzschild black hole and a non-rotating boson star, comparing the outcomes. Our model also incorporates realistic interstellar medium properties, ranging from hot ionized gas to dense molecular clouds. The observational quantity we use to discriminate between the type of compact object is the X-ray flux computed through Bremsstrahlung radiation~\citep{rybicki_radiative_2004}. As we show below, boson star accretion can produce X-ray luminosities 
in the order of $10^{27} \ \text{to} \ 10^{41} \ \text{erg}\,\text{cm}^{-2}\,\text{s}^{-1}$, potentially observable by Chandra. We argue that X-ray observations could 
help discriminating the nature of the dark companions in binary systems like Gaia BH1 and probe the possibility of boson stars as 
alternatives to black holes.

This paper is organized as follows: Section II introduces our boson star model and analyzes spherical accretion of a  polytropic fluid onto black holes and boson stars. In Section III we proceed with the Bondi-Michel calculation for different classes of interstellar mediums (ISMs). Section IV discusses the emissivity and luminosity calculations for the two possible  dark central object candidates we consider in Gaia BH1. Finally, we present our conclusions in Section V. Throughout the paper, the metric signature is $(-,+,+,+)$, and we use natural units $c=G=\hbar=1$, unless stated otherwise. 


\section{Bondi-Michel accretion onto exotic compact objects}
\subsection{Boson star model}

The equations of motion of our boson star model are derived using the action principle. For boson stars, the action must describe Einstein's gravity minimally coupled to a complex, massive, and self-interacting scalar field $\Phi$ (with complex conjugate $\Phi^*$), 
\begin{equation}
    S=\int \sqrt{-g}\left[\frac{R}{16\pi}-\frac{1}{2}\left(\Phi^*_{,\mu}\Phi^{,\mu}+V(|\Phi|^2)\right)\right]d^4x,
\label{eq.1}
\end{equation}
where $V(|\Phi|^2)$ is a self-interacting potential that depends on the magnitude of the scalar field and is invariant under $U(1)$ transformations in the complex plane.
To obtain the field equations, we vary the action with respect to the metric tensor, yielding Einstein's equations,
\begin{equation}
    R_{\mu\nu}-\frac{1}{2}g_{\mu\nu}R=8\pi T_{\mu\nu},
\label{eq.Einstein}
\end{equation}
with the momentum-energy tensor $T_{\mu\nu}$ given by
\begin{equation}
    T_{\mu\nu}=\Phi^*_{,(\mu}\Phi^{}_{,\nu)}-\frac{1}{2}\left[\Phi^*_{,\alpha}\Phi^{,\alpha}+V(|\Phi|^2)\right].
\end{equation}
In the previous equations $R_{\mu\nu}$ is the Ricci tensor, $R$ is its trace, $g_{\mu\nu}$ is the metric, and $g$ its determinant. In addition, the notation `$\Phi^{,\mu}$' indicates partial differentiation with respect to the coordinates $x^{\mu}$. Varying the action for the complex scalar field yields the Klein-Gordon equation
\begin{equation}
    \square\Phi=\frac{dV}{d|\Phi|^2}\Phi.
\label{eq.KleinGordon}
\end{equation}
For our work, we choose the simplest possible boson star model, namely, symmetric static configurations. The metric can be written in Schwarzschild-like coordinates as,
\begin{equation}
    ds^2=-\alpha^2(r)dt^2+a^2(r)dr^2+r^2d\Omega^2,
\label{eq.metric}
\end{equation}
which is expressed in terms of two real metric functions $\alpha(r)$ and $a(r)$, which only depend on the coordinate $r$. This radius is called areal radius since spheres of constant $r$ have a surface area $4\pi r^2$. 
The above line element allows us to define a more general mass function in analogy with the Schwarzschild metric [$a^2=(1-2M/r)^{-1}$]. Thus, we can define the so-called Arnowitt-Deser-Misner (ADM) mass of the spacetime as
\begin{equation}
    M(r,t)=\frac{r}{2}\left(1-\frac{1}{a^2(r,t)}\right).
\end{equation}
This equation measures the total mass contained in a sphere of radius $r$ at a time $t$. 
To construct a physically meaningful model of a boson star, we must ensure a stable and smooth gravitational field in spacetime. To surpass Derrick-type obstructions, we relax the requirement that the scalar field be time-independent while maintaining a time-dependent gravitational field.
Therefore, we assume a harmonic ansatz for the scalar field
\begin{equation}
    \Phi(r,t)=\phi(r)e^{-i\omega t},
\end{equation}
where $\omega$ is the positive field frequency. The time dependence disappears in the stress-energy tensor since it depends only on the absolute value of the field and its gradients.

For this study we consider the simplest case that admits localized solutions, namely a free-field potential given by
\begin{equation}
    V(|\Phi|^2)=m^2|\Phi|^2,
\end{equation}
where $m$ can be interpreted as the bare mass of the field theory. The equilibrium equations to build such ``mini-boson stars'' are obtained by substituting the metric (\ref{eq.metric}) and the harmonic ansatz into the Einstein [Eq.~(\ref{eq.Einstein})] and Klein-Gordon equations [Eq.~(\ref{eq.KleinGordon})]. This results in the following three first-order partial differential equations (see e.g.~\cite{liebling_dynamical_2012}):
\begin{align}
    \partial_ra&=\frac{a}{2}\biggl\{-\frac{a^2-1}{r}+4\pi r\left[\left(\frac{\omega^2}{\alpha^2}+m^2\right)a^2\phi_0^2+\Phi^2\right]  \biggr\}, \label{eq.System1}\\ 
    \partial_r\alpha&=\frac{\alpha}{2}\biggl\{\frac{a^2-1}{r}+4\pi r\left[\left(\frac{\omega^2}{\alpha^2}+m^2\right)a^2\phi_0^2+\Phi^2\right]  \biggr\}, \label{eq.System2}\\ 
    \partial_r\Phi&=-\{1+a^2-4\pi r^2a^2m^2\phi_0^2\}\frac{\Phi}{r}-\left(\frac{\omega^2}{\alpha^2}-m^2\right)\phi_0a^2.
\label{eq.System3}
\end{align}
To obtain a physically meaningful solution, appropriate boundary conditions must be imposed. Regularity at the origin requires
\begin{equation}
    \phi_0(0)=\phi_c,\hspace{0.3cm} \Phi(0)=0, \hspace{0.3cm}a(0)=1.
\end{equation}
On the other hand, asymptotic flatness demands
\begin{align}
    &\lim_{r\to\infty}\phi_0(r)=0,\\  
    &\lim_{r\to\infty}\alpha(r)=\lim_{r\to\infty}\frac{1}{a(r)}.
\end{align}
The system of equations (\ref{eq.System1})-(\ref{eq.System3}) is solved with a shooting method by integrating from $r=0$ towards the outer boundary $r=r_{\rm out}$. For a given central value of the field $\phi_c$, the eigenvalue $\omega$ is adjusted to satisfy the boundary conditions.
For a fixed central field value $\phi_c$, different configurations arise, characterized by distinct effective radii and masses, corresponding to discrete eigenvalues $\omega^{(n)}$. The integer $n$ represents the number of nodes in $\phi_0$ (as $n$ increases, so does the number of nodes). The configuration without nodes corresponds to the ground state of the boson star, while those with nodes represent excited states.

\begin{figure*}[ht]
    \centering
    \begin{minipage}{0.5\textwidth}
        \centering
        \includegraphics[width=\linewidth]{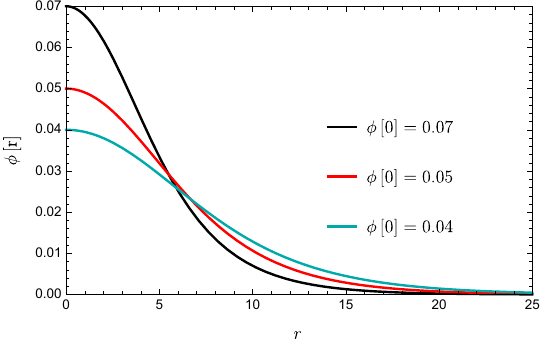}
    \end{minipage}\hfill
    \begin{minipage}{0.5\textwidth}
        \centering
        \includegraphics[width=\linewidth]{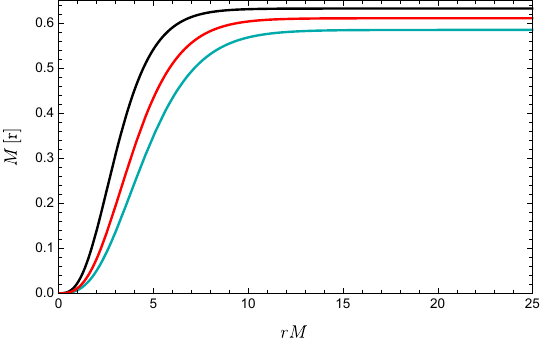}
    \end{minipage}
    \centering
    \begin{minipage}{0.5\textwidth}
        \centering
        \includegraphics[width=\linewidth]{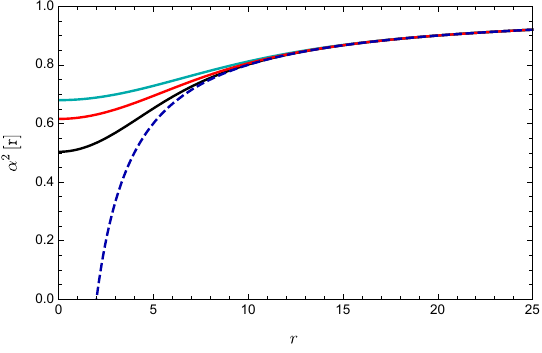}
    \end{minipage}\hfill
    \begin{minipage}{0.5\textwidth}
        \centering
        \includegraphics[width=\linewidth]{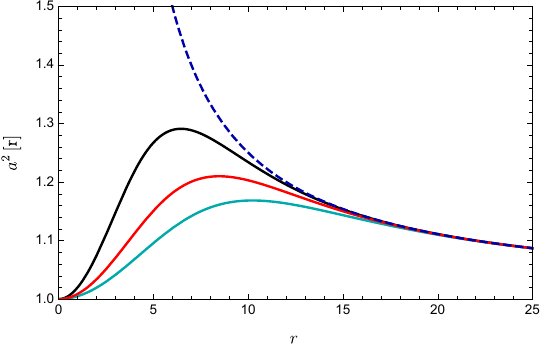}
    \end{minipage}
    \caption{Radial profiles of spacetime metric quantities for three representative mini-boson star solutions   corresponding to various central scalar field values   ($\phi(0)=0.07,0.05,0.04$). The radial coordinate is normalized to the mass of the boson star. \textit{Top left:} Field distributions. Higher central values lead to more compact field distributions, decaying more rapidly with radius. \textit{Top right:}  ADM masses. \textit{Bottom left:} Lapse function $\alpha^2(r)$. Lower central values are attained for larger central field amplitudes. \textit{Bottom right:} Metric component $a^2(r)$. 
    The dashed curves in the bottom panels represent the Schwarzschild black hole solution. The absence of a horizon in boson stars allows the metric to remain smooth across all radii.}
    \label{fig: Boson Metric}
\end{figure*}

Our mini-boson star initial data is reported in Table~\ref{table:1} and displayed in Fig.~\ref{fig: Boson Metric}. In particular, this figure shows different metric and scalar field quantities of mini boson stars in the ground state, where the solutions are computed for the various central field values of Table~\ref{table:1}.
For higher values of the central field, the metric tends to be more asymptotically flat, as the central pressure is smaller, leading to reduced spacetime curvature. As the central pressure increases, stable solutions correspond to larger masses and smaller radii, implying an increase in stellar compactness. However, beyond a critical central field value, the mass of boson stars decreases. Stable mini-boson stars in the ground state can reach a maximum mass that corresponds to $M=0.633M^2_{\rm Pl}/m$, where $M_{\rm Pl}$ is the Planck's mass~\citep{liebling_dynamical_2012}. Before reaching this critical point, increasing $\phi_c$
results in narrower field profiles and larger lapse functions $\alpha$, as the mass increases. 

The dashed lines in the bottom panels of Fig.~\ref{fig: Boson Metric} correspond to the Schwarzschild black hole case. The differences between the metric potentials of boson stars and black holes are more pronounced as we approach the center of these objects (asymptotically the two spacetimes are identical). The most important difference is the lack of an event horizon in the boson star case, which allows to compute the metric for all values of the radius. As we show below, this has important consequences in the accretion patterns onto these compact objects.

\begin{table}
\caption{Boson star parameters for each model considered in this work and represented in Fig.~\ref{fig: Boson Metric}.}              
\label{table:1}      
\centering                                      
\begin{tabular}{c c c }       
\hline                        
$\phi_c$ & $\omega(m/M_{\rm Pl})$ & $M(M_{\rm Pl}^2/m)$  \\    
\hline                                   
    $0.07$ & $0.86256$ & $0.6319$  \\     
    $0.05$ & $0.89524$ & $0.6103$  \\
    $0.04$ & $0.91347$ & $0.5844$  \\
\hline                                             
\end{tabular}
\end{table}

\subsection{Bondi-Michel equations}

To study spherical accretion flows onto a compact object, care must be taken with the characteristics and properties of the flow itself. The boundary conditions imposed at spatial infinity and near the compact object influence the flow dynamics. In this work, we assume that the generalization of the relativistic model maintains the same conditions as the Newtonian Bondi model~\citep{Bondi:1952}. Specifically, we consider a fluid characterized by a fixed pressure, $P_{\infty}$, and density $\rho_\infty$, at large distances. We will assume a polytropic fluid governed by the equation of state $p(\rho)=k\rho^{\Gamma}$, where the pressure is given in terms of the polytropic constant $k$, the polytropic index $\Gamma$ and the density $\rho$. 
For our analysis, we consider the case of a monoatomic ideal gas with $\Gamma=5/3$ at non-relativistic temperatures. 

The governing equations for the spherical accretion of matter onto a nonrotating compact object in general relativity were derived by~\cite{Michel:1972}. Since the mass density and the energy-momentum tensor are divergence-free, the existence of conserved quantities is ensured. Those lead to the mass accretion rate equation and to the Bernoulli equation:
\begin{align}
    \dot{M}&=4\pi\alpha\sqrt{\gamma_{rr}}r^2\rho u^r=\text{const}.  \label{eq:mass_accretion}\\
    -h\mathcal{W}&=h u_t=h\alpha W=\text{const}. \label{eq:bernoulli}
\end{align}
where stationarity and spherical symmetry are assumed. Here, $W=\sqrt{1+\gamma_{rr}u^2}$ is the Lorentz factor and $\gamma$ is the determinant of the spatial metric, given by $\gamma_{ij}=\text{diag}(a^2,r^2,r^2\sin\theta)$. The specific enthalpy $h$ is expressed as
\begin{equation}
    h=1+\left(1+\frac{1}{\Gamma-1}\right) K\rho^{\Gamma-1}\,.
    \label{eq:enthaply}
\end{equation}
Differentiating equations \eqref{eq:mass_accretion} and \eqref{eq:bernoulli}, we obtain the following relation as in \cite{meliani_gr-amrvac_2016}:
\begin{align}
    \left[\left( \frac{d\ln{(\alpha\sqrt{\gamma})}}{d\ln{r}}\right)c_s^2-\frac{\frac{d\alpha^2}{d\ln{r}}+u^2\frac{d(\alpha a)^2}{d\ln{r}}}{2\left(\alpha^2+\alpha^2a^2u^2\right)} \right]& \nonumber \\
    =-\frac{d\ln{u^r}}{d\ln{r}}&\left[c_s^2-\frac{\alpha^2a^2u^2}{\alpha^2+\alpha^2a^2u^2}\right].
    \label{eq:18}
\end{align}
The sound speed $c_s$ is given by $c^2_s=\Gamma p/\rho h$. The last equation gives the conditions at the sonic point, where the flow velocity matches the sound speed. By requiring that both sides of this equation vanish at the sonic point we obtain, from the right-hand-side, the following relation between the sound speed and the radial velocity,
\begin{equation}
    c_{s,c}^2=\frac{a^2u^2_c}{1+a^2u^2_c}\,.
\label{BSsound}
\end{equation}
The left-hand side of Eq.~(\ref{eq:18}) yields the radial velocity
\begin{equation}
    u^2_c=\frac{\frac{d\ln{\alpha}}{d\ln{r}}}{\gamma_{rr}\left(\frac{d\ln{\sqrt{\gamma/a^2}}}{d\ln{r}}\right)}.
\label{BSspeed}
\end{equation}
We note that subindex $c$ in the last two equations indicate that the corresponding quantities are computed at the critical (sonic) point.

Let us now rederive expressions for the mass accretion rate and the Bernoulli equation, in a way similar to  \cite{chakrabarti_solutions_1996}, starting with the 4-velocity $u$, which can be written in terms of the Eulerian velocity $V$ by the transformation 
\begin{equation}
    u^r=V\mathcal{W}=\frac{(-g_{tt})^{1/2}V}{(1+g_{tt}g_{rr}V^2)^{1/2}
    }\,.
    \label{radial_velocity}
\end{equation}
The accretion quantities have been written for a general spherically symmetric metric. This allows us to describe both compact objects  (boson stars and black holes) in a unified framework. In practice, we will express the conserved quantities in terms of the boson star metric only so that to study the black hole case, we simply need to substitute the $\alpha$ and $a$ boson star metric functions with the corresponding Schwarzschild black hole metric functions. 

By expressing the rest-mass density in terms of the sound speed and using Eq.~\eqref{radial_velocity}, we define,
\begin{equation}
    \dot{\mathcal{M}}=\left(\frac{c_s^2}{1-nc_s^2}\right)^{n}\frac{V}{(1-\alpha^2a^2V^2)^{1/2}}r^2\alpha^2 a.
    \label{eq.Mass}
\end{equation}
Similarly, for the Bernoulli energy, we obtain,
\begin{equation}
    E_{\text{Bern}}=\left(\frac{1}{1-nc_s^2}\right)\frac{1}{(1-a^2\alpha^2V^2)^{1/2}}\alpha, \label{eq.Ber}
\end{equation}
where $n=1/(\Gamma-1)$ and $\dot{\mathcal{M}}\propto k^n\dot{M}$.

\begin{figure}[t!]
    \centering
    \begin{minipage}{0.47\textwidth}
        \centering
        \includegraphics[width=\linewidth]{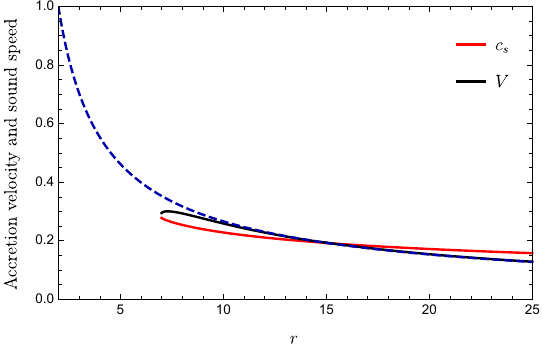}
    \end{minipage}\hfill
    \begin{minipage}{0.47\textwidth}
        \centering
        \includegraphics[width=\linewidth]{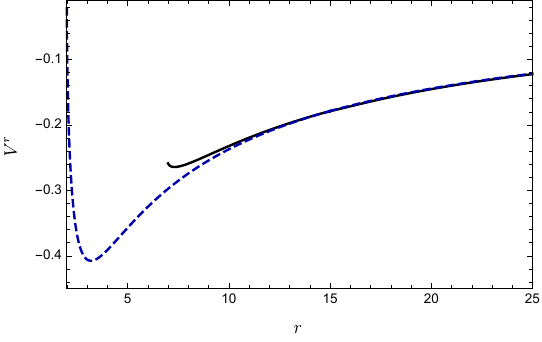}
    \end{minipage}
    \centering
    \caption{\textit{Top:} Fluid velocity and sound speed for a polytropic fluid accreting onto a boson star with mass $M=0.6319M_{\rm Pl}^2/m$. The solution corresponds to a critical radius of $r_c=15$, where there is a change from a supersonic regime to a subsonic one. \textit{Bottom:} Radial component of the fluid velocity. The dashed lines in both panels correspond to the black hole case.}
    \label{fig: Theoretical Accretion}
\end{figure}

\begin{figure}[t!]
    \centering
    \begin{minipage}{0.47\textwidth}
        \centering
        \includegraphics[width=\linewidth]{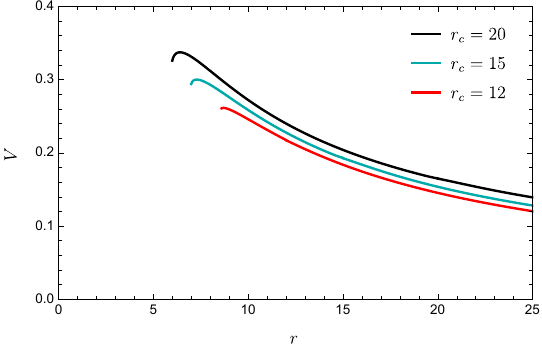}
    \end{minipage}\hfill
    \begin{minipage}{0.47\textwidth}
        \centering
        \includegraphics[width=\linewidth]{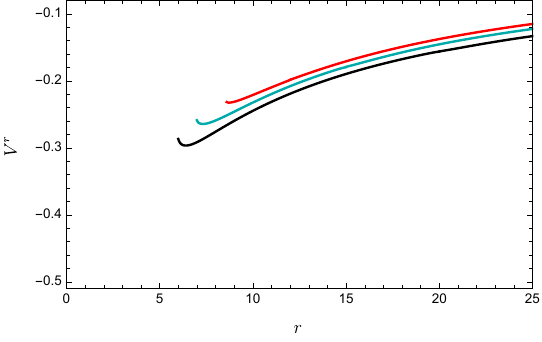}
    \end{minipage}
    \centering
    \caption{Accretion fluid velocity (top) and radial velocity (bottom) for different critical radii. The radius where the gravitational force matches the pressure of the fluid is smaller for bigger critical radii, as the fluid reaches deeper  regions of the boson star.}
    \label{fig: Accretion_cr}
\end{figure}

Using the Newton-Raphson method, we solve equations (\ref{eq.Mass}) and (\ref{eq.Ber}) successively both inward and outward from the sonic point. The results are presented in Figure \ref{fig: Theoretical Accretion}, where we show the accretion velocities for the boson star and black hole cases for a critical radius of $r_c=15$. The fluid remains subsonic at infinity and up to the critical point. Beyond the critical radius, as the fluid moves toward the center of the boson star—where the scalar field is stronger—it transitions into the supersonic regime. The conditions at the critical point $r_c$ are determined by equations (\ref{BSsound}) and (\ref{BSspeed}). 
As the fluid approaches the center, the gravitational force gradually decreases after reaching its maximum.  While gravity is decreasing, the pressure of the fluid begins to counterbalance gravity, which leads to a decrease in the mass accretion rate and, consequently, the velocity of the fluid. For the Bondi-Michel model, it vanishes at a radius of $r=7$. At this point, the outward force exerted by the fluid matches the inward gravitational force. The matching point of pressure and gravity will be closer to the star for a bigger critical radius, as shown in Figure \ref{fig: Accretion_cr}. For a bigger critical radius, the accretion velocity will increase since the sonic point is reached far from the boson star, and the gravitational force will be stronger for bigger critical points. We can compare these results from those by \cite{meliani_gr-amrvac_2016} who studied the accretion parameters for different boson star masses. The boson star model with a smaller ADM mass has a larger gravitational influence when the accreted matter approaches its center because the scalar field potential profile is wider and exerts a stronger gravitational force. 
Far from the boson star the accretion pattern will be similar to the black hole case, as shown in Fig.~\ref{fig: Theoretical Accretion}. Therefore, our results are compatible with those from \cite{meliani_gr-amrvac_2016}, where the accretion dynamics for the two compact objects is indistinguishable after $r=10$.

For a spherically symmetric black hole, the $g_{tt}$ component of the metric decreases towards the black hole, vanishing at the event horizon. The fluid that reaches the horizon will fall into the black hole, eventually becoming causally disconnected from an asymptotic observer. In contrast, in the boson star case there is no event horizon behind which matter can ``disappear". We thus assume that beyond the regime of validity of the accretion solution, matter accumulates in the gravitational potential well produced by the boson star and transitions to a fluid configuration in hydrostatic equilibrium. From now on, we call this region the {\it interior solution} and the domain of validity of the Bondi-Michel model the {\it exterior solution}. In the upcoming sections, we will consider a boson star with a central field of $\phi_c=0.07$ (see Table~\ref{table:1} for details). This corresponds to a compact mini-boson star with a mass lower than the maximum limit of stability. In this way, the accreted fluid can increase the mass of the boson star during the accretion process without triggering the star's instability. 

\subsection{Interior solution}
\begin{figure}[t!]
    \centering
    \begin{minipage}{0.45\textwidth}
        \centering
        \includegraphics[width=\linewidth]{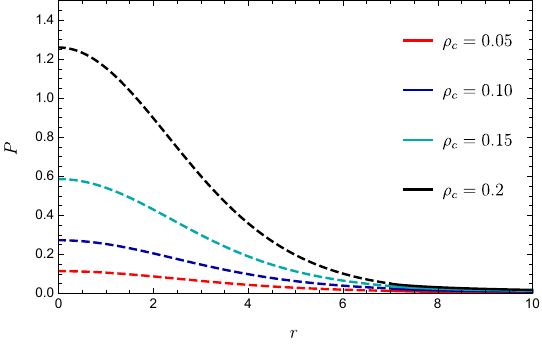}
    \end{minipage}\hfill
    \begin{minipage}{0.45\textwidth}
        \centering
        \includegraphics[width=\linewidth]{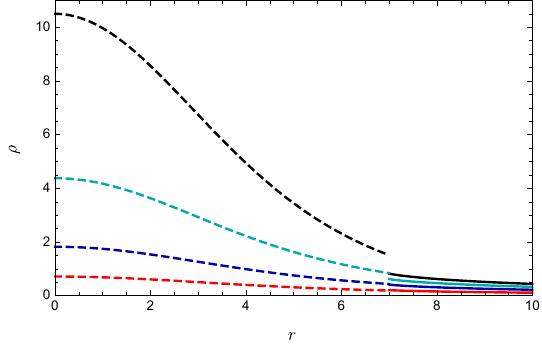}
    \end{minipage}
    \centering
    \caption{\textit{Top:} Polytropic fluid pressure (top) and density (bottom) for the interior and exterior solutions for various values of the critical density and for a fixed critical radius of $r_c=15$. Each critical density solution corresponds to a different value of the polytropic constant in the exterior solution, namely $k_e=[0.2, 0.1, 0.05, 0.025]$ for $\rho_c=[0.05, 0.10, 0.15, 0.20]$. The transition between the exterior and the interior solution happens at $r=5.2$, where pressure is always but density may not be so. All numbers are in natural units.}
    \label{fig: Pressure}
\end{figure}

As a starting point to obtain the interior solution we must derive an equation to express the hydrostatic equilibrium. This involves calculating the momentum energy tensor for a perfect fluid, 
\begin{equation}
    T^{\mu\nu}=(e+p)u^\mu u^\nu+pg^{\mu\nu},
\end{equation}
where $e$ is the total mass-energy density and $u^{\mu}$ the fluid four-velocity. The equilibrium equations will be a solution of  
\begin{align}
    \nabla_\mu(\rho u^\mu)=0,\\
    \nabla_\mu T^{\mu\nu}=0.
\end{align}
Projecting the momentum-energy conservation into the direction orthogonal to the fluid four-velocity using $h^{\mu\nu}=g^{\mu\nu}+u^\mu u^\nu$, the general relativistic Euler equation becomes
\begin{equation}
    (e+p)a_\mu=-h_\mu^\nu\partial_\nu p,
\end{equation}
where $a_\mu=u^\nu\nabla_\nu u_\mu$ is the fluid four-acceleration. For a spherically symmetric and stationary spacetime, $u^\mu=(1,0,0,0)$ and the pressure equilibrium equation reduces to \citep[see][]{rezzolla_brief_2013} 
\begin{equation}
    \frac{dp}{dr}=-(e+p)\frac{d\ln\alpha}{dr}.
    \label{eq:Euler}
\end{equation}
The total energy density $e$ can be written in terms of the internal energy $\epsilon$ and the rest-mass density $\rho$ in the following way,
\begin{equation}
    e=\rho(1+\epsilon)=\rho+\frac{k\rho^\Gamma}{\Gamma-1},
    \label{eq:equilibrium}
\end{equation}
with 
\begin{equation}
    \epsilon=\frac{k\rho^{\Gamma-1}}{\Gamma-1},
\end{equation}
valid for a polytropic equation of state. Thus, Eq.~(\ref{eq:Euler}) can be written in terms of the system's enthalpy. As the fluid is isentropic (the system entropy must remain constant), $(\partial h/\partial p)_S=1/\rho$. This means that the enthalpy can be rewritten as 
\begin{equation}
    h=\frac{e+p}{(dh/dp)^{-1}}.
\end{equation}
We substitute the last equation into Eq.~(\ref{eq:equilibrium}) to obtain the equilibrium equation in differential form 
\begin{equation}
    \frac{d}{dr}\ln{h}=-\frac{d}{dr}\ln{\alpha}, 
\end{equation}
After integrating, we obtain
\begin{equation}
    h(r)\alpha(r)=h(r_0)\alpha(r_0)=\text{const}.
\end{equation}
where $r_0$ can be any point, for example, the center of the star or the boundary between the interior and exterior solutions. We choose $r_0=r_b$ as the point at the boundary between the two solutions. With this choice, the only free variable is the polytropic constant. Notice that, since in general we can expect entropy to change, the value of $k$ is different in the exterior and interior solutions; therefore, to avoid confusion, we will recast the polytropic constant in the interior solution as $k_i$. 
By considering a polytropic fluid $p=k\rho^\Gamma$ and the expression for the enthalpy in Eq.~(\ref{eq:enthaply}), the last equation can be written in terms of the pressure as 
\begin{equation}
    k_i^{1/\Gamma}p^\frac{\Gamma-1}{\Gamma}
    =\left[\left(1+k_i^{1/\Gamma}p_0^\frac{\Gamma-1}{\Gamma}\left(\frac{1}{\Gamma-1}+1\right)\right)\frac{\alpha(r_0)}{\alpha(r)}-1\right]\left(\frac{1}{\Gamma-1}+1\right)^{-1}.
    \label{eq:press}
\end{equation}
Since the value of the polytropic constant is different for the interior and the exterior solution, there is a discontinuity in the density at the point where both solutions meet. However, the pressure from the two solutions must match at that point, ensuring continuity. This behavior is seen in Figure \ref{fig: Pressure}, where the pressure and the density are obtained from Eq.~(\ref{eq:press}). Each critical density corresponds to a value of the polytropic constant; the smaller the polytropic constant, the bigger the pressure and the density in the exterior solution. It can be expected that most of the fluid mass is contained in the interior solution.

\section{Accretion in the interstellar medium}

\begin{table*}[ht!]
\caption{Characteristics of the different interstellar mediums present in the Milky Way Galaxy. Given the density and the typical temperature of the ISM, the sound speed at infinity is calculated through equation (\ref{eq.Temp}). }              
\label{table:2}      
\centering                                      
\begin{tabular}{c c c c}       
\hline                        
ISM& $\rho$ (particles$\,\,$$\text{cm}^{-3}$) & $T$$(K)$& $c_{s,\infty}$ (km/s) \\    
\hline                                   
    HIM & $10^{-4}-10^{-2}$ & $10^6$& $117.29$ \\     
    WIM & $0.2-0.5$ & $8000$ & $10.49$\\
    WNM & $0.2-0.5$ & $6000-10000$ & $9.09-11.73$\\
    CNM & $20-50$ & $50-100$ & $0.83-1.17$\\
    Molecular Clouds & $10^2-10^5$ & 0.5844 & $0.37-0.83$ \\
\hline                                             
\end{tabular}
\end{table*}

The accretion solution discussed in the previous section provides a theoretical understanding of how fluids behave when accreting onto compact objects under the assumption of spherical symmetry. The differences between boson stars and black holes become apparent as we approach the center of these objects. However, no differences are expected at large distances. To gain a deeper insight into the accretion process, we should examine how these objects behave in real astrophysical environments. In an actual physical situation compact objects coexist with a background density, and accretion patterns depend strongly on the rest-mass density and temperature of the underlying interstellar medium. These factors directly influence the accretion rate and, consequently, the emitted luminosity. 

\begin{table*}[ht!]
\caption{Accretion parameters regarding the ISMs at the critical point. The conserved quantities, such as the Bernoulli energy and the mass accretion rate at the critical point are calculated using the parameters at infinity (see table \ref{table:2}).}              
\label{table:3}      
\centering                                      
\begin{tabular}{c c c c c}       
\hline                        
ISM& $r_c/M$ & $c_{s,c}$& $\rho_c(\text{particles}\,\,\text{cm}^{-3})$  &$\dot{M}(\text{M}_\odot\,\, \text{yr}^{-1})$ \\    
\hline                                   
    HIM & $1918$ & $1.61\times10^{-2}$ & $7.045-704.5$ & $9.03\times(10^{-23}-10^{-21})$ \\     
    WIM & $21447$ & $4.83\times10^{-3}$ & $(0.53-1.31)\times10^6$&$(2.52-6.31)\times10^{-16}$\\
    WNM & $(2.48-1.92)\times10^{4}$ & $(4.49-5.10)\times10^{-3}$ & $(0.45-1.63)\times10^6$ & $(1.81-9.71)\times10^{-16}$ \\
    CNM & $(2.71-1.92)\times10^{5}$ & $(1.36-1.61)\times10^{-3}$ & $(1.41-5.93)\times10^9$&$(0.18-1.28)\times10^{-10}$\\
    Molecular Clouds & $(6.06-2.71)\times10^5$ & $(9.08-13.57)\times10^{-4}$ & $1.84\times10^{10}-3.96\times10^{13}$ & $2.55\times10^{-10}-2.85\times10^{-6}$ \\
\hline                                             
\end{tabular}
\end{table*}

To obtain the Bondi-Michel solution in realistic environments, we first examine typical ISM environments in our galaxy, summarized in Table \ref{table:2} \citep[see][for a detailed analysis on the density of the ISMs]{saintonge_cold_2022,ferriere_interstellar_2001}. These include the hot ionized medium (HIM), which is less dense and hotter than the warm ionized medium (WIM) and warm neutral medium (WNM), and the cold neutral medium (CNM). We also consider the properties of molecular clouds, which is the coldest environment but the most dense one. The sound speed at infinity, $c_{s,\infty}$ is related to the temperature of the ISM. Assuming the ISM behaves as a monoatomic hydrogen fluid, we compute the sound speed using
\begin{equation}
    c_{s,\infty}=\left(  \Gamma\frac{k_B}{m_p}T  \right)^{1/2},
    \label{eq.Temp}
\end{equation}
where $k_B$ is the Boltzmann constant and $m_p$ the proton mass. To solve the accretion problem, we start by calculating the sound speed and rest-mass density at infinity and at the critical point using the Bernoulli conserved energy equation,
\begin{equation}
    (h\mathcal{W})_\infty=(h\mathcal{W})_c.
\end{equation}
The class of spherically symmetric spacetimes we are considering are asymptotically flat, meaning $\alpha^2=a^2=1$. We also assume the fluid is asymptotically at rest, thus $\mathcal{W}_\infty=1$. At the critical point, we use the relation between equations (\ref{BSsound}) and (\ref{BSspeed}) to write the Lorentz factor as 
\begin{equation}
    \mathcal{W}_c=\alpha_c\sqrt{1+\frac{c_{s,c}^2}{1-c_{s,c}^2}}.
\end{equation}
Thus, Bernoulli equation simplifies to
\begin{equation}
    h_\infty=h_c\alpha_c\sqrt{1+\frac{c_{s,c}^2}{1-c_{s,c}^2}}.
\end{equation}
Expressing the enthalpy in terms of the polytropic fluid density [see Eq.~(\ref{eq:enthaply})], we obtain
\begin{equation}
    1+\frac{c_{s,\infty}^2}{\Gamma-1-c_{s,\infty}^2}=\left(1+\frac{c_{s,c}^2}{\Gamma-1-c_{s,c}^2}\right)\alpha_c\sqrt{1+\frac{c_{s,c}^2}{1-c_{s,c}^2}},
    \label{eq:cc}
\end{equation}
where we have used the equation
\begin{equation}
    \Gamma k\rho^{\Gamma-1}=\frac{(\Gamma-1)c_s^2}{\Gamma-1-c_s^2},
\end{equation}
to express the density in terms of the sound speed. This allows us to relate the fluid parameters at infinity and at the critical point and thus calculate the mass accretion rate using the Bondi-Michel solution. Since the polytropic constant is the same throughout the domain, we obtain
\begin{equation}
    \rho_{c}^{1-\Gamma}=\frac{c_{s,\infty}^2}{c_{s,c}^2}\left(\frac{\Gamma-1-c_{s,c}^2}{\Gamma-1-c_{s,\infty}^2}\right)\rho_\infty^{1-\Gamma}.
    \label{eq:rhoc}
\end{equation}
Using Eq.~(\ref{eq:cc}) we can obtain the sound speed at the critical point, and with Eq.~(\ref{eq:rhoc}), we obtain the critical density. Table~\ref{table:3} reports these quantities for each ISM, along with the critical radius and the mass accretion rate, which are calculated by assuming that at infinity, the accretion process in boson stars follows the same process as for black holes. From \citet{rezzolla_brief_2013} the critical radius is given (for $\Gamma=5/3$) by
\begin{equation}
    r_c=\frac{3}{4}\frac{M}{c_{s,\infty}}\,.
\end{equation}
Since the mass accretion rate is a conserved quantity, we can express it with the parameters at infinity as in \citep{rezzolla_brief_2013,aguayo-ortiz_spherical_2021}:
\begin{equation}
   \dot{M}=4\pi\lambda_B(GM)^2\frac{\rho_\infty}{c_{s,\infty}^3}, 
\end{equation}
with $\lambda_B=1/4$. Here, we have explicitly introduced the gravitational constant $G$ and the mass of the compact object $M$. 

It is possible to conduct a quick survey of the stability of our boson star models in the conditions of each ISM we consider. For instance, with the mass accretion rates from Table~\ref{table:3} we can study the time needed to increase the star mass by $1\%$ of its total mass. For a boson star of $10M_{\odot}$, the time required to increase its mass would be around $2.75\times10^{18-16}$ years for the HIM and $1.37\times10^7-1.94\times10^{6}$ years for the CNM.  These timescales are comparable to or exceed the age of the universe ($\approx10^{10}$ years), implying the star can remain stable for extended periods of time. However, this depends on factors such as the mass increase and the star's mass. As the mass of the boson star increases, the time necessary to increase the mass by $1\%$ will decrease. Furthermore, if a $10M_\odot$ boson star is accreting from molecular clouds, the time to increase its mass by $1\%$ is around $9.70\times10^5-86.90$ years, which would increase the boson star mass at a higher rate and surpass the stability limit.

\begin{figure}[t!]
    \centering
    \begin{minipage}{0.48\textwidth}
        \centering
        \includegraphics[width=\linewidth]{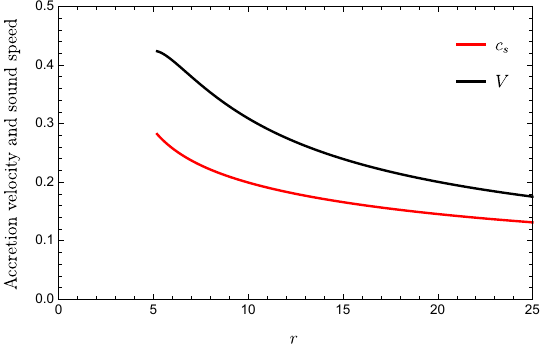}
    \end{minipage}\hfill
    \begin{minipage}{0.48\textwidth}
        \centering
        \includegraphics[width=\linewidth]{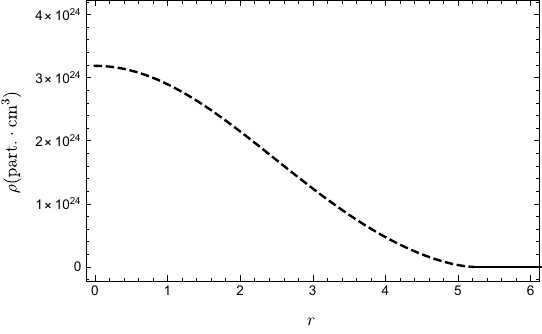}
    \end{minipage}\hfill
    \begin{minipage}{0.48\textwidth}
        \centering
        \includegraphics[width=\linewidth]{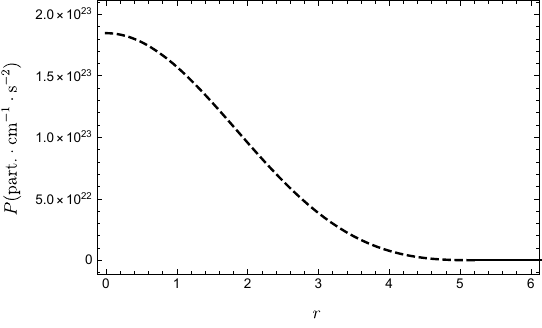}
    \end{minipage}
    \centering
    \caption{\textit{Top panel:} Accretion flow of a polytropic fluid onto a boson star for the HIM. The critical radius is placed at $r_c=1918$ and the transition between the exterior and interior solution occurs at $r=5.2$, where the star's gravitational force matches the pressure exerted by the fluid. \textit{Middle and bottom panels:} Corresponding radial profiles of the density and of the pressure, respectively. The dashed lines in these two panels indicates the interior solution if we consider a boson star as a central object for tha parameters of Gaia BH1, where the polytropic constant is $k_i=2.676\times10^{-18}$ (see Section $4.1$ and Table~\ref{table:4} for details).}
    \label{fig:HIM}
\end{figure}

Finally, the accretion solution in each ISM is calculated through the mass accretion rate and the Bernoulli equation. For the specific parameters of the HIM the accretion velocity and sound speed are displayed in Figure \ref{fig:HIM}. The boundary between the exterior and the interior solution is placed at $r=5.2$. Once we have obtained the spherically symmetric accretion solution onto a boson star in the ISM, we are in a position to study its associated X-ray emission.

\section{Bremsstrahlung emission}
To quantitatively analyze the X-ray emission arising from accretion onto compact objects, we begin by integrating the total Bremsstrahlung radiation produced during the accretion process. This free-free emission originates from collisions between electrons and ions, where the acceleration of charged particles in a Coulomb field results in the emission of radiation, particularly at X-ray wavelengths. As in previous sections, we assume the accreting medium is composed solely of hydrogen and is electrically neutral. Consequently, the number densities of electrons and ions are equal on average. 
In the calculation of the Bremsstrahlung emissivity from a fluid accreted onto a boson star, we start with the expression provided by \cite{rybicki_radiative_2004},
\begin{equation}
    \epsilon_{BR}=1.7\times10^{-27}T^{1/2}\rho^2 Z^2 g_B\ \frac{\text{erg}}{\text{cm}^3\cdot \text{s}}.
\end{equation}
Here, $\rho$ is the rest-mass density from the accretion solution, the atomic number of ions is $Z=1$ for a monoatomic fluid and the Gaunt factor, $g_\text{B}$ is taken as 1.2. The temperature of the fluid can be written in terms of the density and pressure through Eq.~(\ref{eq.Temp}). Using the definition of the sound speed we can rewrite the equation for the emissivity as
\begin{equation}
 \epsilon_{BR}=1.85\times10^{-31}\rho^2\left(\frac{p}{\rho}\right)^{1/2}. 
\end{equation}
The total X-ray luminosity from the system is obtained by integrating the emissivity across the volume \citep[see][]{olivares_general_2023},
\begin{equation}
 L_{BR}=\int \epsilon_{BR}W \sqrt{\gamma} d^3x.
\end{equation}
Due to the assumption of spherical symmetry, the volume element becomes $d^3x=drd\theta d\phi$, and with the spatial metric determinant $\gamma$ and the Lorentz factor $W$ we obtain
\begin{equation}
 L_{BR}=4\pi M^3\left(\frac{G}{c^2}\right)^3\int \epsilon_{BR}a^2r^2\sqrt{1+a^2u^2}  dr. 
 \label{luminosity}
\end{equation}
The prefactors outside the integral are constants necessary to calculate the luminosity in physical units.

\subsection{The central dark object of Gaia observations}

As discussed in the Introduction, the binary system Gaia BH1 reported by the Gaia astrometric mission is composed of a solar-type star of mass $M=0.93M_\odot$ orbiting around a central dark object of mass $M_{\text{DO}}=9.62M_\odot$. 
The lack of X-ray or radio-wave observations from this binary makes hard to identify the true nature of the dark object. While the Gaia mission \citep{gaia_collaboration_gaia_2023,el-badry_sun-like_2022} identifies it as a black hole, the dynamics leading to the formation of this system challenge this interpretation. On the other hand, the alternative explanation of~\cite{pombo_sun-like_2023} hypothesizes the possibility that boson stars might be the dark compact objects of some observed binary systems. In this section we calculate the X-ray luminosity (cf.~Eq.~(\ref{luminosity}) from Gaia BH1 to test the two hypotheses -- black hole vs boson star -- and understand the main differences. Our computations will be done for a Hot Ionized Medium and we will use the specifications of the Chandra X-ray observatory to evaluate the possibility to observe X-rays due to the accretion onto the dark central object. We will assume that the central object captures matter from the companion star through spherically symmetric wind accretion. For this model, the mass accretion rate is given by \citep[see][]{frank_accretion_2002}, 
\begin{equation}
    \dot{M}= \frac{G^2M_{\text{DO}}^2\dot{M}_\text{wind}}{v_\text{wind}^4 a^2},
    \label{eq:MBH1}
\end{equation}
where $\dot{M}_\text{wind}$ is the star wind mass-loss rate, $v_\text{wind}$ is the wind velocity, and $a$ the semi-major axis. Typical G-dwarf stars have wind mass-loss rates approximately around $10^{-14} 
\ M_\odot \, \text{yr}^{-1}$ and the mass accretion rate given by Eq.~(\ref{eq:MBH1}) is $\dot{M}\sim10^{-18}\ M_\odot \,\text{yr}^{-1}$ 

In Figure \ref{fig: Pressure} we display different accretion solutions depending on the polytropic constant chosen for the interior solution. Different solutions have different masses and correspond to different boson star accretion histories. We can expect that more trapped mass means more luminosity. To calculate the values for the polytropic constant in the interior solution, we integrate the density inside the star. This increase in the mass should correspond to the total mass of the fluid that has been accreted for a given time. This means
\begin{equation}
  M_t= 4\pi m_p \int_0^{5.2} \rho_{\text{in}}\sqrt{\gamma_{rr}} r^2dr=\dot{M}\Delta t,
\end{equation}
where the interior density $\rho_{\text{in}}$ is given by Eq.~(\ref{eq:press}) by considering the polytropic equation of state. 

Finally, to obtain the integration limits for the luminosity we check Chandra's resolution. For Chandra, the angular resolution is $0.5$ arcsec and the sensitivity is $4\times10^{-15} \text{erg}\, \text{cm}^{-2}\, \text{s}^{-1}$ in $10^5\ \text{s}$. The distance at which Chandra can measure radiation is then given by
\begin{equation}
 \tan  \left(\frac{\delta}{2}\right)=\frac{d}{2D},
\end{equation}
where $\delta$ is the angular resolution, $d$ is the integration limit and $D$ the distance from the telescope to the source. The Gaia BH1 binary is at a distance of $477$ pc from Earth, therefore, the integration limit is $d=1.15625\times10^{-3}$ pc. The distance can be expressed in terms of the mass by calculating the gravitational radius for the source
\begin{equation}
    r_g=\frac{GM}{c^2}.
\end{equation}
The gravitational radius of the Gaia binary source with a mass of $M=9.62M_\odot$ is $r_g=2.52\times10^9M$. 
Table~\ref{table:4} reports the values of the polytropic constant $k_i$ for different accreting times $\Delta t$, along with the Bremsstrahlung luminosity and the X-ray radiation flux. As the boson star accretes more mass, the value of the polytropic constant decreases, and the luminosity increases. For the X-ray emission to be detected by the Chandra telescope, the flux of the  radiation detected must be higher than the telescope's sensitivity, namely $4\times10^{-15} \text{
erg}\, \text{cm}^{-2}\, \text{s}^{-1}$. Therefore, we consider a spherical surface area $A$ with a radius equal to the distance from the telescope to the BH1 binary ($477$ pc). If the boson star had been accreting for 10 years (first row in Table~\ref{table:4}), the X-ray radiation should not have been detected by Chandra as the flux is lower than the telescope's sensitivity. However, at around $10^2$ years of accretion (second row) the luminosity would become visible by the detector, and X-ray emission from this source could be expected. This is because, contrary to the black hole case, the fluid can enter inside the (horizonless) boson star while remaining causally connected to distant observers, and the emissivity in this region is higher for the interior solution than for the exterior one. It is worth noting that the computation for the black hole case was reported by Gaia~\citep{gaia_collaboration_gaia_2023}, yielding low X-ray luminosities, around  $5.676\times10^{27}$ erg$\,$$\text{s}^{-1}$. The corresponding flux would be of the order of $2.50\times10^{-16}\,$erg$\,$$\text{cm}^{-2}$$\,$$\text{s}^{-1}$, which is below Chandra's sensitivity. 

\begin{table}[t!]
\caption{Dependence of the Bremsstrahlung luminosity and X-ray radiation flux for the Gaia BH1 system on the accretion time, considering a boson star as a central dark object and that the system evolves in a hot ionized medium. The polytropic constant $k_i$ for the interior accretion solution depends on the fluid mass accreted by the compact object. For a Chandra sensitivity of $4\times10^{-15} \text{
erg}\, \text{cm}^{-2}\, \text{s}^{-1}$ an observable flux would require an accreting time higher than $10^2$ years.}              
\label{table:4}      
\centering                                      
\begin{tabular}{c c c c}       
\hline                        
 $\Delta t $& $k_i$ & $L$ & $L/A$  \\    
 (years)      &  & (erg$\,\,$$\text{s}^{-1}$) & (erg$\,\,$$\text{cm}^{-2}$$\,\,$$\text{s}^{-1}$)\\      
\hline                                   
    $10$ & $5.765\times10^{-15}$ & $6.047\times10^{27}$ & $2.221\times10^{-16}$\\     
    $10^2$ & $1.242\times10^{-15}$ & $6.047\times10^{29}$ & $2.221\times10^{-14}$\\
    $10^4$ & $5.765\times10^{-17}$ & $6.047\times10^{33}$ & $2.221\times10^{-10}$\\
    $10^6$ & $2.676\times10^{-18}$  & $6.047\times10^{37}$ & $2.221\times10^{-6}$ \\
    $10^8$ & $1.242\times10^{-19}$ & $6.047\times10^{41}$ & $2.221\times10^{-2}$ \\
\hline                                             
\end{tabular}
\end{table}

We need to be conservative in these results since the star is orbiting the dark object with a semi-major axis distance of 1.4 AU, corresponding to $6.79\times10^{-6}$ pc. This means that when we measure the radiation from the accretion onto the dark object, we will also measure the radiation coming from the star itself since the integration limit is larger than their orbital separation. Furthermore, the object is accreting mass at a rate of $10^{-18}\ \text{M}_\odot\, \text{yr}^{-1}$. This means that the environment surrounding the central dark object has intermediate properties between the HIM and the WIM (see Table~\ref{table:3}). In principle, the density of the external region can be rescaled to reproduce the same critical parameters, such as the sound speed and critical radius, while adopting the appropriate density for the actual medium. However, in the case of a boson star, the dominant contribution to the luminosity arises from the interior solution. As such, the assumption of an HIM environment remains valid and does not significantly affect the emitted radiation.


\section{Conclusions}
We have conducted a comparative analysis between the X-ray emission produced by black holes and boson stars by assuming spherically symmetric accretion of a polytropic test fluid onto both types of compact objects. The discovery of the Gaia binary BH1, and in particular the lack of X-ray emission in this system, has been the driving motivation to study the nature of the central dark object. Using the Bondi-Michel accretion framework for a polytropic equation of state describing a monoatomic ideal gas, we have derived the mass accretion rate and Bernoulli equation for both a static Schwarzschild spacetime and for a boson star metric.

Our results show that accretion onto boson stars is distinct from black hole accretion in the interior region. While both objects exhibit similar behavior at large radii, significant deviations emerge near the center of the boson star, due to the absence of an event horizon and the presence of a smooth scalar field. The fluid slows down as pressure gradients begin to counteract gravity, contrasting with the monotonic infall seen in black holes. These differences are most apparent near the inner radii of the boson star and have important implications for observables like X-ray emission.

A quantitative analysis of the X-ray emission resulting from spherical accretion  was performed. 
Using the Bremsstrahlung emissivity, we derived the total luminosity expected from such accretion processes and compared the detectability of the resulting X-ray flux using Chandra's observational constraints.
We applied this framework to the binary system Gaia BH1. While standard interpretations suggest the dark compact object could be a black hole, the lack of detectable X-ray or radio-wave emission raises questions about this classification. Our analysis shows that for spherical wind accretion at realistic mass-loss rates, a black hole would produce X-ray luminosities below the detection threshold of Chandra. In contrast, the boson star hypothesis predicts detectable X-ray fluxes after an accretion period of approximately $10^2$ years or more due to the presence of matter inside the object and higher emissivity in the interior solution.

The results reported in this work suggest that X-ray observations can be an important diagnostic tool to make a distinction between black holes and alternative compact objects such as boson stars. While current instruments may not detect radiation from short-term accretion events, future observations or more prolonged accretion histories could reveal X-ray signals that favor the existence of exotic compact objects. Further observations and the development of state-of-the-art telescopes are essential to constrain the true nature of dark companions in binary systems such as Gaia BH1. On the theoretical side, future extensions of the  research reported in this paper should consider more complex spacetimes by incorporating angular momentum, exploring alternative theoretical models for exotic compact objects, and considering different accretion frameworks that may affect X-ray emission in such binary systems.

\begin{acknowledgements}
This work is supported by CIDMA under the FCT Multi-Annual Financing Program for R\&D Units and by the European Horizon Europe staff exchange (SE) programme HORIZON-MSCA-2021-SE-01 Grant No. NewFunFiCO-101086251. HO is supported by the Individual CEEC program - 5th edition funded by the Portuguese Foundation for Science and Technology. HO further acknowledges support from the projects PTDC/FIS-AST/3041/2020, CERN/FIS-PAR/0024/2021 and 2022.04560.PTDC. JAF acknowledges support from the Spanish Agencia Estatal de Investigaci\'on (grant PID2021-125485NB-C21) funded by MCIN/AEI/10.13039/501100011033 and ERDF A way of making Europe and from the Generalitat Valenciana (grant CIPROM/2022/49). AO is partially supported by FCT under the contract CERN/FIS-PAR/0037/2021.
\end{acknowledgements}

\bibliographystyle{aa} 
\bibliography{Bibliography.bib} 

\begin{thebibliography}{33}
\expandafter\ifx\csname natexlab\endcsname\relax\def\natexlab#1{#1}\fi

\bibitem[{Aguayo-Ortiz {et~al.}(2021)Aguayo-Ortiz, Tejeda, Sarbach, \& López-Cámara}]{aguayo-ortiz_spherical_2021}
Aguayo-Ortiz, A., Tejeda, E., Sarbach, O., \& López-Cámara, D. 2021, Monthly Notices of the Royal Astronomical Society, 504, 5039

\bibitem[{{Bezares} \& {Sanchis-Gual}(2024)}]{2024arXiv240604901B}
{Bezares}, M. \& {Sanchis-Gual}, N. 2024, arXiv e-prints, arXiv:2406.04901

\bibitem[{{Bondi}(1952)}]{Bondi:1952}
{Bondi}, H. 1952, \mnras, 112, 195

\bibitem[{Bustillo {et~al.}(2021)Bustillo, Sanchis-Gual, Torres-Forné, Font, Vajpeyi, Smith, Herdeiro, Radu, \& Leong}]{bustillo_gw190521_2021}
Bustillo, J.~C., Sanchis-Gual, N., Torres-Forné, A., {et~al.} 2021, Physical Review Letters, 126, 081101, arXiv:2009.05376 [gr-qc]

\bibitem[{{Calder{\'o}n Bustillo} {et~al.}(2023){Calder{\'o}n Bustillo}, {Sanchis-Gual}, {Leong}, {Chandra}, {Torres-Forn{\'e}}, {Font}, {Herdeiro}, {Radu}, {Wong}, \& {Li}}]{bustillo_searching_2023}
{Calder{\'o}n Bustillo}, J., {Sanchis-Gual}, N., {Leong}, S. H.~W., {et~al.} 2023, \prd, 108, 123020

\bibitem[{Carloni \& Rosa(2019)}]{carloni_derricks_2019}
Carloni, S. \& Rosa, J.~L. 2019, Physical Review D, 100, 025014, arXiv:1906.00702 [gr-qc]

\bibitem[{Chakrabarti(1996)}]{chakrabarti_solutions_1996}
Chakrabarti, S.~K. 1996, The Astrophysical Journal, 471, 237, publisher: IOP Publishing

\bibitem[{Chaverra {et~al.}(2016)Chaverra, Mach, \& Sarbach}]{chaverra_michel_2016}
Chaverra, E., Mach, P., \& Sarbach, O. 2016, Classical and Quantum Gravity, 33, 105016, publisher: IOP Publishing

\bibitem[{Derrick(1964)}]{derrick_comments_1964}
Derrick, G.~H. 1964, J. Math. Phys., 5, 1252

\bibitem[{{El-Badry} {et~al.}(2023){El-Badry}, {Rix}, {Quataert}, {Howard}, {Isaacson}, {Fuller}, {Hawkins}, {Breivik}, {Wong}, {Rodriguez}, {Conroy}, {Shahaf}, {Mazeh}, {Arenou}, {Burdge}, {Bashi}, {Faigler}, {Weisz}, {Seeburger}, {Almada Monter}, \& {Wojno}}]{2023MNRAS.518.1057E}
{El-Badry}, K., {Rix}, H.-W., {Quataert}, E., {et~al.} 2023, \mnras, 518, 1057

\bibitem[{El-Badry {et~al.}(2022)El-Badry, Rix, Quataert, Howard, Isaacson, Fuller, Hawkins, Breivik, Wong, Rodriguez, Conroy, Shahaf, Mazeh, Arenou, Burdge, Bashi, Faigler, Weisz, Seeburger, Almada Monter, \& Wojno}]{el-badry_sun-like_2022}
El-Badry, K., Rix, H.-W., Quataert, E., {et~al.} 2022, Monthly Notices of the Royal Astronomical Society, 518, 1057

\bibitem[{Ferrière(2001)}]{ferriere_interstellar_2001}
Ferrière, K.~M. 2001, Reviews of Modern Physics, 73, 1031, publisher: American Physical Society

\bibitem[{Frank {et~al.}(2002)Frank, King, \& Raine}]{frank_accretion_2002}
Frank, J., King, A., \& Raine, D. 2002, Accretion {Power} in {Astrophysics}, 3rd edn. (Cambridge: Cambridge University Press)

\bibitem[{{Gaia Collaboration} {et~al.}(2023){Gaia Collaboration}, Arenou, Babusiaux, Barstow, Faigler, Jorissen, Kervella, Mazeh, Mowlavi, Panuzzo, Sahlmann, Shahaf, Sozzetti, Bauchet, Damerdji, Gavras, Giacobbe, Gosset, Halbwachs, Holl, Lattanzi, Leclerc, Morel, Pourbaix, Re~Fiorentin, Sadowski, Ségransan, Siopis, Teyssier, Zwitter, Planquart, Brown, Vallenari, Prusti, De~Bruijne, Biermann, Creevey, Ducourant, Evans, Eyer, Guerra, Hutton, Jordi, Klioner, Lammers, Lindegren, Luri, Mignard, Panem, Randich, Sartoretti, Soubiran, Tanga, Walton, Bailer-Jones, Bastian, Drimmel, Jansen, Katz, Van~Leeuwen, Bakker, Cacciari, Castañeda, De~Angeli, Fabricius, Fouesneau, Frémat, Galluccio, Guerrier, Heiter, Masana, Messineo, Nicolas, Nienartowicz, Pailler, Riclet, Roux, Seabroke, Sordo, Thévenin, Gracia-Abril, Portell, Altmann, Andrae, Audard, Bellas-Velidis, Benson, Berthier, Blomme, Burgess, Busonero, Busso, Cánovas, Carry, Cellino, Cheek, Clementini, Davidson, De~Teodoro, Nuñez~Campos, Delchambre, Dell’Oro,
  Esquej, Fernández-Hernández, Fraile, Garabato, García-Lario, Haigron, Hambly, Harrison, Hernández, Hestroffer, Hodgkin, Janßen, Jevardat De~Fombelle, Jordan, Krone-Martins, Lanzafame, Löffler, Marchal, Marrese, Moitinho, Muinonen, Osborne, Pancino, Pauwels, Recio-Blanco, Reylé, Riello, Rimoldini, Roegiers, Rybizki, Sarro, Smith, Utrilla, Van~Leeuwen, Abbas, Ábrahám, Abreu~Aramburu, Aerts, Aguado, Ajaj, Aldea-Montero, Altavilla, Álvarez, Alves, Anders, Anderson, Anglada~Varela, Antoja, Baines, Baker, Balaguer-Núñez, Balbinot, Balog, Barache, Barbato, Barros, Bartolomé, Bassilana, Becciani, Bellazzini, Berihuete, Bernet, Bertone, Bianchi, Binnenfeld, Blanco-Cuaresma, Blazere, Boch, Bombrun, Bossini, Bouquillon, Bragaglia, Bramante, Breedt, Bressan, Brouillet, Brugaletta, Bucciarelli, Burlacu, Butkevich, Buzzi, Caffau, Cancelliere, Cantat-Gaudin, Carballo, Carlucci, Carnerero, Carrasco, Casamiquela, Castellani, Castro-Ginard, Chaoul, Charlot, Chemin, Chiaramida, Chiavassa, Chornay, Comoretto,
  Contursi, Cooper, Cornez, Cowell, Crifo, Cropper, Crosta, Crowley, Dafonte, Dapergolas, David, De~Laverny, De~Luise, De~March, De~Ridder, De~Souza, De~Torres, Del~Peloso, Del~Pozo, Delbo, Delgado, Delisle, Demouchy, Dharmawardena, Diakite, Diener, Distefano, Dolding, Enke, Fabre, Fabrizio, Fedorets, Fernique, Figueras, Fournier, Fouron, Fragkoudi, Gai, Garcia-Gutierrez, Garcia-Reinaldos, García-Torres, Garofalo, Gavel, Gerlach, Geyer, Gilmore, Girona, Giuffrida, Gomel, Gomez, González-Núñez, González-Santamaría, González-Vidal, Granvik, Guillout, Guiraud, Gutiérrez-Sánchez, Guy, Hatzidimitriou, Hauser, Haywood, Helmer, Helmi, Sarmiento, Hidalgo, Hilger, Hładczuk, Hobbs, Holland, Huckle, Jardine, Jasniewicz, Jean-Antoine~Piccolo, Jiménez-Arranz, Juaristi~Campillo, Julbe, Karbevska, Khanna, Kordopatis, Korn, Kóspál, Kostrzewa-Rutkowska, Kruszyńska, Kun, Laizeau, Lambert, Lanza, Lasne, Le~Campion, Lebreton, Lebzelter, Leccia, Lecoeur-Taibi, Liao, Licata, Lindstrøm, Lister, Livanou, Lobel, Lorca,
  Loup, Madrero~Pardo, Magdaleno~Romeo, Managau, Mann, Manteiga, Marchant, Marconi, Marcos, Marcos~Santos, Marín~Pina, Marinoni, Marocco, Marshall, Martin~Polo, Martín-Fleitas, Marton, Mary, Masip, Massari, Mastrobuono-Battisti, McMillan, Messina, Michalik, Millar, Mints, Molina, Molinaro, Molnár, Monari, Monguió, Montegriffo, Montero, Mor, Mora, Morbidelli, Morris, Muraveva, Murphy, Musella, Nagy, Noval, Ocaña, Ogden, Ordenovic, Osinde, Pagani, Pagano, Palaversa, Palicio, Pallas-Quintela, Panahi, Payne-Wardenaar, Peñalosa~Esteller, Penttilä, Pichon, Piersimoni, Pineau, Plachy, Plum, Poggio, Prša, Pulone, Racero, Ragaini, Rainer, Raiteri, Ramos, Ramos-Lerate, Regibo, Richards, Rios~Diaz, Ripepi, Riva, Rix, Rixon, Robichon, Robin, Robin, Roelens, Rogues, Rohrbasser, Romero-Gómez, Rowell, Royer, Ruz~Mieres, Rybicki, Sáez~Núñez, Sagristà~Sellés, Salguero, Samaras, Sanchez~Gimenez, Sanna, Santoveña, Sarasso, Schultheis, Sciacca, Segol, Segovia, Semeux, Siddiqui, Siebert, Siltala, Silvelo, Slezak,
  Slezak, Smart, Snaith, Solano, Solitro, Souami, Souchay, Spagna, Spina, Spoto, Steele, Steidelmüller, Stephenson, Süveges, Surdej, Szabados, Szegedi-Elek, Taris, Taylor, Teixeira, Tolomei, Tonello, Torra, Torra, Torralba~Elipe, Trabucchi, Tsounis, Turon, Ulla, Unger, Vaillant, Van~Dillen, Van~Reeven, Vanel, Vecchiato, Viala, Vicente, Voutsinas, Weiler, Wevers, Wyrzykowski, Yoldas, Yvard, Zhao, Zorec, \& Zucker}]{gaia_collaboration_gaia_2023}
{Gaia Collaboration}, Arenou, F., Babusiaux, C., {et~al.} 2023, Astronomy \& Astrophysics, 674, A34

\bibitem[{{Gaia Collaboration} {et~al.}(2016){Gaia Collaboration}, Prusti, De~Bruijne, Brown, Vallenari, Babusiaux, Bailer-Jones, Bastian, Biermann, Evans, Eyer, Jansen, Jordi, Klioner, Lammers, Lindegren, Luri, Mignard, Milligan, Panem, Poinsignon, Pourbaix, Randich, Sarri, Sartoretti, Siddiqui, Soubiran, Valette, Van~Leeuwen, Walton, Aerts, Arenou, Cropper, Drimmel, Høg, Katz, Lattanzi, O’Mullane, Grebel, Holland, Huc, Passot, Bramante, Cacciari, Castañeda, Chaoul, Cheek, De~Angeli, Fabricius, Guerra, Hernández, Jean-Antoine-Piccolo, Masana, Messineo, Mowlavi, Nienartowicz, Ordóñez-Blanco, Panuzzo, Portell, Richards, Riello, Seabroke, Tanga, Thévenin, Torra, Els, Gracia-Abril, Comoretto, Garcia-Reinaldos, Lock, Mercier, Altmann, Andrae, Astraatmadja, Bellas-Velidis, Benson, Berthier, Blomme, Busso, Carry, Cellino, Clementini, Cowell, Creevey, Cuypers, Davidson, De~Ridder, De~Torres, Delchambre, Dell’Oro, Ducourant, Frémat, García-Torres, Gosset, Halbwachs, Hambly, Harrison, Hauser, Hestroffer,
  Hodgkin, Huckle, Hutton, Jasniewicz, Jordan, Kontizas, Korn, Lanzafame, Manteiga, Moitinho, Muinonen, Osinde, Pancino, Pauwels, Petit, Recio-Blanco, Robin, Sarro, Siopis, Smith, Smith, Sozzetti, Thuillot, Van~Reeven, Viala, Abbas, Abreu~Aramburu, Accart, Aguado, Allan, Allasia, Altavilla, Álvarez, Alves, Anderson, Andrei, Anglada~Varela, Antiche, Antoja, Antón, Arcay, Atzei, Ayache, Bach, Baker, Balaguer-Núñez, Barache, Barata, Barbier, Barblan, Baroni, Barrado Y~Navascués, Barros, Barstow, Becciani, Bellazzini, Bellei, Bello~García, Belokurov, Bendjoya, Berihuete, Bianchi, Bienaymé, Billebaud, Blagorodnova, Blanco-Cuaresma, Boch, Bombrun, Borrachero, Bouquillon, Bourda, Bouy, Bragaglia, Breddels, Brouillet, Brüsemeister, Bucciarelli, Budnik, Burgess, Burgon, Burlacu, Busonero, Buzzi, Caffau, Cambras, Campbell, Cancelliere, Cantat-Gaudin, Carlucci, Carrasco, Castellani, Charlot, Charnas, Charvet, Chassat, Chiavassa, Clotet, Cocozza, Collins, Collins, Costigan, Crifo, Cross, Crosta, Crowley, Dafonte,
  Damerdji, Dapergolas, David, David, De~Cat, De~Felice, De~Laverny, De~Luise, De~March, De~Martino, De~Souza, Debosscher, Del~Pozo, Delbo, Delgado, Delgado, Di~Marco, Di~Matteo, Diakite, Distefano, Dolding, Dos~Anjos, Drazinos, Durán, Dzigan, Ecale, Edvardsson, Enke, Erdmann, Escolar, Espina, Evans, Eynard~Bontemps, Fabre, Fabrizio, Faigler, Falcão, Farràs~Casas, Faye, Federici, Fedorets, Fernández-Hernández, Fernique, Fienga, Figueras, Filippi, Findeisen, Fonti, Fouesneau, Fraile, Fraser, Fuchs, Furnell, Gai, Galleti, Galluccio, Garabato, García-Sedano, Garé, Garofalo, Garralda, Gavras, Gerssen, Geyer, Gilmore, Girona, Giuffrida, Gomes, González-Marcos, González-Núñez, González-Vidal, Granvik, Guerrier, Guillout, Guiraud, Gúrpide, Gutiérrez-Sánchez, Guy, Haigron, Hatzidimitriou, Haywood, Heiter, Helmi, Hobbs, Hofmann, Holl, Holland, Hunt, Hypki, Icardi, Irwin, Jevardat De~Fombelle, Jofré, Jonker, Jorissen, Julbe, Karampelas, Kochoska, Kohley, Kolenberg, Kontizas, Koposov, Kordopatis,
  Koubsky, Kowalczyk, Krone-Martins, Kudryashova, Kull, Bachchan, Lacoste-Seris, Lanza, Lavigne, Le~Poncin-Lafitte, Lebreton, Lebzelter, Leccia, Leclerc, Lecoeur-Taibi, Lemaitre, Lenhardt, Leroux, Liao, Licata, Lindstrøm, Lister, Livanou, Lobel, Löffler, López, Lopez-Lozano, Lorenz, Loureiro, MacDonald, Magalhães~Fernandes, Managau, Mann, Mantelet, Marchal, Marchant, Marconi, Marie, Marinoni, Marrese, Marschalkó, Marshall, Martín-Fleitas, Martino, Mary, Matijevič, Mazeh, McMillan, Messina, Mestre, Michalik, Millar, Miranda, Molina, Molinaro, Molinaro, Molnár, Moniez, Montegriffo, Monteiro, Mor, Mora, Morbidelli, Morel, Morgenthaler, Morley, Morris, Mulone, Muraveva, Musella, Narbonne, Nelemans, Nicastro, Noval, Ordénovic, Ordieres-Meré, Osborne, Pagani, Pagano, Pailler, Palacin, Palaversa, Parsons, Paulsen, Pecoraro, Pedrosa, Pentikäinen, Pereira, Pichon, Piersimoni, Pineau, Plachy, Plum, Poujoulet, Prša, Pulone, Ragaini, Rago, Rambaux, Ramos-Lerate, Ranalli, Rauw, Read, Regibo, Renk, Reylé,
  Ribeiro, Rimoldini, Ripepi, Riva, Rixon, Roelens, Romero-Gómez, Rowell, Royer, Rudolph, Ruiz-Dern, Sadowski, Sagristà~Sellés, Sahlmann, Salgado, Salguero, Sarasso, Savietto, Schnorhk, Schultheis, Sciacca, Segol, Segovia, Segransan, Serpell, Shih, Smareglia, Smart, Smith, Solano, Solitro, Sordo, Soria~Nieto, Souchay, Spagna, Spoto, Stampa, Steele, Steidelmüller, Stephenson, Stoev, Suess, Süveges, Surdej, Szabados, Szegedi-Elek, Tapiador, Taris, Tauran, Taylor, Teixeira, Terrett, Tingley, Trager, Turon, Ulla, Utrilla, Valentini, Van~Elteren, Van~Hemelryck, Van~Leeuwen, Varadi, Vecchiato, Veljanoski, Via, Vicente, Vogt, Voss, Votruba, Voutsinas, Walmsley, Weiler, Weingrill, Werner, Wevers, Whitehead, Wyrzykowski, Yoldas, Žerjal, Zucker, Zurbach, Zwitter, Alecu, Allen, Allende~Prieto, Amorim, Anglada-Escudé, Arsenijevic, Azaz, Balm, Beck, Bernstein, Bigot, Bijaoui, Blasco, Bonfigli, Bono, Boudreault, Bressan, Brown, Brunet, Bunclark, Buonanno, Butkevich, Carret, Carrion, Chemin, Chéreau, Corcione,
  Darmigny, De~Boer, De~Teodoro, De~Zeeuw, Delle~Luche, Domingues, Dubath, Fodor, Frézouls, Fries, Fustes, Fyfe, Gallardo, Gallegos, Gardiol, Gebran, Gomboc, Gómez, Grux, Gueguen, Heyrovsky, Hoar, Iannicola, Isasi~Parache, Janotto, Joliet, Jonckheere, Keil, Kim, Klagyivik, Klar, Knude, Kochukhov, Kolka, Kos, Kutka, Lainey, LeBouquin, Liu, Loreggia, Makarov, Marseille, Martayan, Martinez-Rubi, Massart, Meynadier, Mignot, Munari, Nguyen, Nordlander, Ocvirk, O’Flaherty, Olias~Sanz, Ortiz, Osorio, Oszkiewicz, Ouzounis, Palmer, Park, Pasquato, Peltzer, Peralta, Péturaud, Pieniluoma, Pigozzi, Poels, Prat, Prod’homme, Raison, Rebordao, Risquez, Rocca-Volmerange, Rosen, Ruiz-Fuertes, Russo, Sembay, Serraller~Vizcaino, Short, Siebert, Silva, Sinachopoulos, Slezak, Soffel, Sosnowska, Straižys, Ter~Linden, Terrell, Theil, Tiede, Troisi, Tsalmantza, Tur, Vaccari, Vachier, Valles, Van~Hamme, Veltz, Virtanen, Wallut, Wichmann, Wilkinson, Ziaeepour, \& Zschocke}]{gaia_collaboration_gaia_2016}
{Gaia Collaboration}, Prusti, T., De~Bruijne, J. H.~J., {et~al.} 2016, Astronomy \& Astrophysics, 595, A1

\bibitem[{Guzman(2006)}]{guzman_accretion_2006}
Guzman, F.~S. 2006, Phys. Rev. D, 73, 021501, \_eprint: gr-qc/0512081

\bibitem[{Herdeiro {et~al.}(2021)Herdeiro, Pombo, Radu, Cunha, \& Sanchis-Gual}]{herdeiro_imitation_2021}
Herdeiro, C.~A., Pombo, A.~M., Radu, E., Cunha, P.~V., \& Sanchis-Gual, N. 2021, Journal of Cosmology and Astroparticle Physics, 2021, 051, publisher: IOP Publishing

\bibitem[{{Liebling} \& {Palenzuela}(2023)}]{liebling_dynamical_2012}
{Liebling}, S.~L. \& {Palenzuela}, C. 2023, Living Reviews in Relativity, 26, 1

\bibitem[{Lora-Clavijo {et~al.}(2013)Lora-Clavijo, Guzmán, \& Cruz-Osorio}]{lora-clavijo_pbh_2013}
Lora-Clavijo, F.~D., Guzmán, F.~S., \& Cruz-Osorio, A. 2013, Journal of Cosmology and Astroparticle Physics, 2013, 015

\bibitem[{Mach {et~al.}(2018)Mach, Piróg, \& Font}]{mach_relativistic_2018}
Mach, P., Piróg, M., \& Font, J.~A. 2018, Classical and Quantum Gravity, 35, 095005, publisher: IOP Publishing

\bibitem[{Malec(1999)}]{malec_fluid_1999}
Malec, E. 1999, Physical Review D, 60, 104043, arXiv:gr-qc/9907028

\bibitem[{Meliani {et~al.}(2016)Meliani, Grandclément, Casse, Vincent, Straub, \& Dauvergne}]{meliani_gr-amrvac_2016}
Meliani, Z., Grandclément, P., Casse, F., {et~al.} 2016, Classical and Quantum Gravity, 33, 155010, publisher: IOP Publishing

\bibitem[{{Michel}(1972)}]{Michel:1972}
{Michel}, F.~C. 1972, \apss, 15, 153

\bibitem[{Olivares {et~al.}(2020)Olivares, Younsi, Fromm, Laurentis, Porth, Mizuno, Falcke, Kramer, \& Rezzolla}]{olivares_how_2020}
Olivares, H., Younsi, Z., Fromm, C.~M., {et~al.} 2020, Monthly Notices of the Royal Astronomical Society, 497, 521, arXiv:1809.08682 [gr-qc]

\bibitem[{Olivares {et~al.}(2023)Olivares, Mościbrodzka, \& Porth}]{olivares_general_2023}
Olivares, H.~R., Mościbrodzka, M.~A., \& Porth, O. 2023, Astronomy \& Astrophysics, 678, A141

\bibitem[{Pombo \& Saltas(2023)}]{pombo_sun-like_2023}
Pombo, A.~M. \& Saltas, I. 2023, Monthly Notices of the Royal Astronomical Society, 524, 4083

\bibitem[{Ressler {et~al.}(2021)Ressler, Quataert, White, \& Blaes}]{ressler_magnetically_2021}
Ressler, S.~M., Quataert, E., White, C.~J., \& Blaes, O. 2021, Monthly Notices of the Royal Astronomical Society, 504, 6076

\bibitem[{Rezzolla \& Zanotti(2013)}]{rezzolla_brief_2013}
Rezzolla, L. \& Zanotti, O. 2013, in Relativistic {Hydrodynamics}, ed. L.~Rezzolla \& O.~Zanotti (Oxford University Press), 0

\bibitem[{Rosa \& Rubiera-Garcia(2022)}]{rosa_shadows_2022}
Rosa, J.~L. \& Rubiera-Garcia, D. 2022, Physical Review D, 106, 084004, arXiv:2204.12949 [gr-qc]

\bibitem[{Rybicki \& Lightman(2004)}]{rybicki_radiative_2004}
Rybicki, G.~B. \& Lightman, A.~P. 2004, Radiative processes in astrophysics (Weinheim, [Germany]: Wiley-VCH Verlag GmbH \& Co. KGaA)

\bibitem[{Saintonge \& Catinella(2022)}]{saintonge_cold_2022}
Saintonge, A. \& Catinella, B. 2022, Annual Review of Astronomy and Astrophysics, 60, 319

\bibitem[{Visinelli(2021)}]{visinelli_boson_2021}
Visinelli, L. 2021, International Journal of Modern Physics D, 30, 2130006, arXiv:2109.05481 [gr-qc]

\bibitem[{Weih {et~al.}(2020)Weih, Olivares, \& Rezzolla}]{weih_two-moment_2020}
Weih, L.~R., Olivares, H., \& Rezzolla, L. 2020, Monthly Notices of the Royal Astronomical Society, 495, 2285

\end{thebibliography}

\end{document}